\documentclass[Physsubmission, Phys]{SciPost}

\hypersetup{
    colorlinks,
    linkcolor={red!50!black},
    citecolor={blue!50!black},
    urlcolor={blue!80!black}
}

\usepackage[bitstream-charter]{mathdesign}
\usepackage{bm}
\usepackage{dsfont}
\newcommand{\I}{\mathrm{i}}
\newcommand{\INT}{\int d^4 x\;}
\newcommand{\Tr}{\mathrm{Tr}}
\newcommand{\Int}{\mathrm{int}}
\newcommand{\g}[1]{\gamma_{#1}}
\newcommand{\su}{SU(2)_{CS}}
\newcommand{\sun}{SU(2)_{CS}^{\mathcal{P}}}
\newcommand{\diag}{\mathrm{diag}}
\DeclareSymbolFont{usualmathcal}{OMS}{cmsy}{m}{n}
\DeclareSymbolFontAlphabet{\mathcal}{usualmathcal}

\begin{document}

\begin{center}{\Large \textbf{
Chiralspin symmetry and confinement\\
}}\end{center}

\begin{center}
Marco Catillo\textsuperscript{1}
\end{center}

\begin{center}
{\bf 1} Institute for Theoretical Physics, ETH Z\"urich, 8093 Z\"urich, Switzerland
\\
* mcatillo@phys.ethz.ch
\end{center}

\begin{center}
\today
\end{center}


\definecolor{palegray}{gray}{0.95}
\begin{center}
\colorbox{palegray}{
  \begin{minipage}{0.95\textwidth}
    \begin{center}
    {\it  XXXIII International (ONLINE) Workshop on High Energy Physics \\“Hard Problems of Hadron Physics:  Non-Perturbative QCD \& Related Quests”}\\
    {\it November 8-12, 2021} \\
    \doi{10.21468/SciPostPhysProc.?}\\
    \end{center}
  \end{minipage}
}
\end{center}

\section*{Abstract}
{\bf
Interesting lattice QCD simulations at high temperature in QCD and particular truncated studies have shown the emergence of an unexpected group symmetry, so called \textit{chiralspin}.
However this is not a symmetry of the QCD action for free quarks, which makes unclear the transition to deconfinement at high temperature in QCD. 
Therefore we try to redefine this group so that is a symmetry of free quark action and it is consistent with the presence of deconfinement in QCD.
}

\vspace{10pt}
\noindent\rule{\textwidth}{1pt}
\tableofcontents\thispagestyle{fancy}
\noindent\rule{\textwidth}{1pt}
\vspace{10pt}

\section{Introduction}
\label{sec:intro}

Recent lattice QCD calculations \cite{Rohrhofer:2017grg,Rohrhofer:2019qwq,Rohrhofer:2019qal,Glozman:2021jlk} have shown that exists a phase in QCD at high temperature where matter  becomes \textit{chiralspin} symmetric (denoted as stringy fluid in \cite{Rohrhofer:2017grg,Rohrhofer:2019qwq,Rohrhofer:2019qal,Glozman:2021jlk}). The \textit{chiralspin} group, or $\su$, is quite peculiar. Indeed, it is not a symmetry of the free quark action, which makes it not so compatible with the regime of deconfinement in QCD. 
However, from the other hand lattice QCD \textit{truncated studies} \cite{Denissenya:2014poa,Denissenya:2014ywa,Denissenya:2015mqa} (where in section \ref{sec:chiralspin} we will explain in what they consist), have pointed out that $\su$ appears together with the emergence of chiral but also axial symmetry. 
The compromise for this, is having $\su$ at $T>T_c$ (with $T_c$ the chiral phase transition temperature), where $U(1)_A$ is approximately restored, but not at too high temperature since QCD goes in the phase of deconfinement, where quarks interact more weakly (quark-gluon plasma). 
Lattice QCD studies therefore found $\su$ as an approximate symmetry in the  range $T_c -3T_c$. 
Nevertheless, the mechanism on how the transition to this \textit{chiralspin} symmetry regime occurs and then vanishes is not completely clear. 
Moreover the fact that from \textit{truncated studies} $\su$ is present together with chiral and axial symmetry but differently from them, $\su$ is not a symmetry of free massless quark action, leads to a veil of mystery on it. 

Therefore in this proceeding we propose to construct a new type of \textit{chiralspin} group in euclidean space-time (in section \ref{sec:chiralspinnew}), which we denote as $\sun$ (we name it  \textit{P-chiralspin} group) that is a symmetry of the free massless quark action and that can possibly explain the \textit{truncated studies}  
results \cite{Rohrhofer:2017grg,Rohrhofer:2019qwq,Rohrhofer:2019qal,Glozman:2021jlk} and consequently solving the issues previously mentioned with $\su$ regarding deconfinement.
For doing this, we study temporal correlators where the space coordinate are kept fixed and then we see that a possible mass degeneracy which could be driven by the presence of a \textit{chiralspin} symmetry can be also perfectly explained by the \textit{P-chiralspin} one (look section \ref{sec:correlators}). 
This, as has been done for $\su$ symmetry, gives also consequences at high temperature QCD, where the presence of \textit{P-chiralspin} can be plausible, even at non-zero chemical potential. 
However lattice studies on this direction are extremely important for having an indication that this hypothesis is correct.
We also give in section \ref{sec:chiralspinnew} a constraint on the gauge field properties in order to have such $\sun$ symmetry in case a gauge interaction is introduced.

\section{Chiralspin group} 
\label{sec:chiralspin}

The \textit{chiralspin} group, or $\su$, is defined in euclidean space-time by the following generators \cite{Glozman:2015qva}, 

\begin{equation}
	\Sigma_n = \{\g{4},\I\g{5}\g{4},-\g{5}\},
	\label{eq:su2cs_generator}
\end{equation}

\noindent
where $\g{4,5}$ are the usual gamma matrices. 
It easy to show that they form an $su(2)$ algebra, because $[\Sigma_n,\Sigma_m ]=  2\I\epsilon_{nmk}\Sigma_k$, $\Sigma_n^{\dagger} = \Sigma_n$ and $\Tr(\Sigma_n) = 0$, for all $n=1,2,3$. 
The $\su$ transformations for quark fields $\psi$ and $\bar{\psi}$ are 

\begin{equation}
	\psi(x)\to\exp(\I\alpha_n\Sigma_n)\psi(x),\qquad
	\bar{\psi}(x)\to\bar{\psi}(x)\g{4}\exp(-\I\alpha_n\Sigma_n)\g{4},
	\label{eq:su2cs_barpsi}
\end{equation}

\noindent
where the 2nd transformation has been taken thinking to the minkowskian version of $\psi$ (namely $\bar{\psi}_M = \psi^{\dagger}_M\g{4}$). 
It is interesting to observe that since $\g{5}$ is one of the generators of $\su$, then $U(1)_A \subset\su$. 
Therefore having $\su$ symmetry implies the axial symmetry as well. 
The transformations (\ref{eq:su2cs_barpsi}), has been used for explaining the large mass degeneracy in the hadron spectrum coming from the \textit{truncated studies} on lattice QCD simulations. Let us remind what these kind of studies are. 
For simplicity we take mesons (for baryons the argument is totally the same) and we start from a generic meson observable $O_{\Gamma}(x) = \bar{\psi}(x)\Gamma\psi(x)$. 
Here $\Gamma$ is a matrix acting on the space of Dirac and flavor (but eventually also color) indices and therefore it specifies the quantum numbers of the meson in consideration. 
We take other 3 observables substituting $\Gamma\to\Gamma\Sigma_n$ ($n=1,2,3$). Then the following correlators 

\begin{equation}
	C_{X}(t) = \sum_{\bm{x}}\langle O_X (x)\bar{O}_X(y)\rangle,
	\label{eq:corr}
\end{equation} 

\noindent
with $X \in\{\Gamma,\Gamma\Sigma_1,\Gamma\Sigma_2,\Gamma\Sigma_3\}$, $y=(\bm{0},0)$ and $x=(\bm{x},t)$, are all connected via $\su$ and they are in general different at zero temperature in QCD. 
For practical purposes, in lattice QCD is convenient to rewrite (\ref{eq:corr}) in terms of the quark propagator $D^{-1}$, inverse of the Dirac operator $D$. In this situation, Eq. (\ref{eq:corr}) becomes 

\begin{equation}
	C_{X}(t) = \sum_{\bm{x}}\langle \Tr(XD^{-1}(x,x))\Tr(\g{4}X^{\dagger}\g{4}D^{-1}(y,y)) -  \Tr(D^{-1}(x,y)\g{4}X^{\dagger}\g{4}D^{-1}(y,x)X)\rangle,
		\label{eq:corr2}
\end{equation}

\noindent
where the first term is called \textit{disconnected} and the second is the \textit{connected} one, while the trace $\Tr(\cdot)$ is over Dirac, flavor and color indices.
The \textit{truncated studies} of Refs. \cite{Denissenya:2014poa,Denissenya:2014ywa,Denissenya:2015mqa} consist in substituting in (\ref{eq:corr2}) the quark propagator with a new one as follow

\begin{equation}
	D^{-1}\to D^{-1}_{(\Lambda)} = D^{-1}-\sum_{\lambda_l : \vert\lambda_l\vert<\Lambda} \frac{1}{\lambda_l}\vert \lambda_l\rangle\langle \lambda_l \vert,
	\label{eq:prop}
\end{equation}

\noindent
where $\lambda_l$ and $\vert \lambda_l\rangle$ are eigenvalues and eigenvectors of $D$ and $\Lambda >0$ is some parameter to be tuned. 
Therefore in (\ref{eq:prop}) the lowest eigenmodes of $D$ are manually removed in $D^{-1}$ and the result is considering the truncated quark propagator $D^{-1}_{(\Lambda)}$.
From the correlators (\ref{eq:corr2}), one can get the hadron masses, exploiting that at large $t$ we have $C_{X}(t) \sim \exp (-m_X t)$. It has been observed that after substitution (\ref{eq:prop}) in (\ref{eq:corr2}) and for $\Lambda$ up to $\sim 180\, MeV$ at least, such exponential decay behavior still persists. We can denote the new correlator as $C_{X}^{(\Lambda)}(t)$ and we therefore have $C_{X}^{(\Lambda)}(t) \sim \exp (-m_X^{(\Lambda)} t)$ for large $t$.
Now, while all $m_X$s for $X \in\{\Gamma,\Gamma\Sigma_1,\Gamma\Sigma_2,\Gamma\Sigma_3\}$ are in general all different, however after removing $\sim 10$ eigenmodes (which corresponds to $\Lambda\sim 65\, MeV$) and restricting on gauge configurations with zero topological charge $Q_{top}=0$, then the masses $m_X^{(\Lambda)}$s get all degenerate. This means that we are in presence of the $\su$ symmetry. Therefore also $U(1)_A\subset\su$ is restored. 
In reality this is nor the only observed thing. There is also a further hadron mass degeneration due to the restoration of chiral symmetry $SU(N_F)_L\times SU(N_F)_R$, which has been 
explained by the group $SU(2N_F)$, that contains $SU(N_F)_L\times SU(N_F)_R \times \su$ as subgroup \cite{Denissenya:2014poa,Denissenya:2014ywa,Denissenya:2015mqa,Glozman:2015qva}. 

This suggests us to speculate that $\su$ should emerge in a regime where at least these conditions are satisfied: 1) Gauge configurations with $Q_{top} = 0$ are dominant; 2) the lowest eigenmodes of $D$ are suppressed; 3) Chiral and axial symmetries emerge.
A physical regime of QCD where at least approximately these conditions are satisfied is at high temperature above chiral phase transition $T_c$, as L. Glozman in \cite{Glozman:2017dfd,Glozman:2016ayd} suggested. Indeed, the lattice results of Refs. \cite{Rohrhofer:2017grg,Rohrhofer:2019qwq,Rohrhofer:2019qal,Glozman:2021jlk} have shown that in the range of temperatures $T_c - 3 T_c$, the $\su$ symmetry appears in hadron correlators. 
However for $T>3T_c$ this symmetry vanishes. The reason is evident. QCD at high temperature approaches to a theory of weakly interacting quarks (deconfinement), but as we explain in the next section $\su$ is not a symmetry of free quark action and therefore not compatible with such regime. Nevertheless, in the range of temperature $T_c-3T_c$ we can assume that quarks are still strongly interacting and therefore the presence of $\su$ is well reasonable.

\section{New chiralspin group definition}
\label{sec:chiralspinnew}

The chiralspin group as has been defined in Eq. (\ref{eq:su2cs_barpsi}), presents some interesting aspects. 
Indeed, in contrast with chiral and axial group, $\su$ is not a symmetry of free massless quark action $S_F = \INT \bar{\psi}(x)\g{\mu}\partial_{\mu}\psi(x)$. 
This fact can be explained writing a general element $U = \exp(\I\alpha_n\Sigma_n)$, with $U\in\su$, as product of three $U(1)$ matrices belonging to the groups $U(1)_A\subset\su$ (generated by $\g{5}$) and $U(1)_4\subset\su$ (generated by $\g{4}$, see (\ref{eq:su2cs_generator})).  
This can always be done for whatever element in $\su$. 
Namely $U = U_A^{\beta_1}U_4^{\beta_2}U_A^{\beta_3}$, where $U_A^{\beta_{1,3}}=\exp(-\I\beta_{1,3}\g{5})\in U(1)_A$ and $U_4^{\beta_2} = \exp(\I\beta_{2}\g{4})\in U(1)_4$. 
Now as shown in Refs. \cite{Catillo:2021rrq,Catillo:2021awx}, while $U(1)_A$ is a symmetry of free massless quark action, $U(1)_4$ is the part of $\su$ which is not a symmetry of $S_F$, because $ \INT \bar{\psi}(x)\g{i}\partial_{i}\psi(x)$, for $i=1,2,3$ is not $U(1)_4$ invariant. 
The problem is now that at first, since $\su$ is not a symmetry of the action of free quarks, then it is not clear from where it comes from. Secondly, if at high temperature QCD looks to approach in the deconfinement then we can ask on why $\su$ shouldn't be compatible with it. Third, we can still ask ourself, if we are really sure that there are not other ways (another \textit{chiralspin} definition) which also can explain the mass degeneration of the \textit{truncated studies}.

Therefore here we will try to redefine $U(1)_4$ and consequently $\su$ in order to make $S_F$ invariant. 
The solution that we came up in Refs.  \cite{Catillo:2021rrq,Catillo:2021awx} exploits the parity transformation for spinors. In formulae we define in substitution of $U(1)_4$ this other group transformations

\begin{equation}
	U(1)_P:\;\psi(x)\to\sum_{n=0}^{\infty}\frac{(\I\alpha)^n}{n!} \psi(x)^{\mathcal{P}^n},\quad
	\bar{\psi}(x)\to\sum_{n=0}^{\infty}\frac{(-\I\alpha)^n}{n!} \bar{\psi}(x)^{\mathcal{P}^n},
	\label{eq:u1p}
\end{equation}

\noindent
where $\psi(x)^{\mathcal{P}^n} = \g{4}^n\psi(\mathcal{P}^n x)$ and 
$\bar{\psi}(x)^{\mathcal{P}^n} = \bar{\psi}(\mathcal{P}^n x)\g{4}^n$ with $\mathcal{P} = \diag(-1,-1,-1,1)$ the parity matrix, so $\mathcal{P}x = (-\bm{x},x_4)$. 
Now, using that $\g{4}^{2k} = \mathds{1}$, 
$\forall k$, we can expand (\ref{eq:u1p}) as 

\begin{equation}
		U(1)_P:\,\psi(x)\to \cos(\alpha)\psi(x)+\I\sin(\alpha)\g{4}\psi(\mathcal{P}x),\quad
		\bar{\psi}(x)\to \cos(\alpha)\bar{\psi}(x)-\I\sin(\alpha)\bar{\psi}(\mathcal{P}x)\g{4}.
		\label{eq:u1p2}	
\end{equation}

As we can see the definition of $U(1)_P$ transformations is pretty similar to $U(1)_4$, with the difference that a parity transformation is applied to the term proportional to $\g{4}$. 
As shown in Ref. \cite{Catillo:2021awx}, $U(1)_P$ is now a symmetry of $S_F$ while $U(1)_4\subset\su$ is not. 
Therefore $U(1)_P$ is more suitable to construct a new type of \textit{chiralspin} group which is a symmetry of the free massless action, that includes the subgroup $U(1)_A$. 
For this aim, we define $\psi_{\pm}(x) = (\psi(x)\pm\psi(\mathcal{P}x))/2$ and $\bar{\psi}_{\pm}(x) = (\bar{\psi}(x)\pm\bar{\psi}(\mathcal{P}x))/2$ and then we introduce the fields 

\begin{equation}
\Psi(x) =\left(
\begin{matrix}
\psi_{+}(x)\\
\psi_{-}(x)
\end{matrix}
\right),\qquad
\bar{\Psi}(x) = \left(\begin{matrix}\bar{\psi}_{+}(x)& \bar{\psi}_{-}(x)\end{matrix}\right).
\label{eq:psi}
\end{equation}

Directly from (\ref{eq:u1p2}), $U(1)_P$ transformations for $\Psi$ and $\bar{\Psi}$ read as  $\Psi(x)\to\exp(\I\alpha(\sigma_3\otimes\g{4}))\Psi(x)$ and 
$\bar{\Psi}(x)\to\bar{\Psi}(x)\g{4}\exp(-\I\alpha(\sigma_3\otimes\g{4}))\g{4}$, where $\sigma_3 = \diag(\mathds{1},-\mathds{1})$ acts in the 2-dimensional space defined in (\ref{eq:psi}). 
The $U(1)_A$ transformations for $\Psi$ and $\bar{\Psi}$ can be obtained in the same way from the transformations of $\psi$ and $\bar{\psi}$. 
We obtain that $\Psi(x)\to\exp(\I\alpha(-\mathds{1}\otimes\g{5}))\Psi(x)$ 
and $\bar{\Psi}(x)\to\bar{\Psi}(x)\g{4}\exp(-\I\alpha(-\mathds{1}\otimes\g{5}))\g{4}$. 
Now taking the generators

\begin{equation}
	\Sigma_n^{\mathcal{P}}=\{\sigma_3\otimes\g{4},\sigma_3\otimes\I\g{5}\g{4},-\mathds{1}\otimes\g{5}\},
	\label{eq:gennew}
\end{equation}

\noindent
where we defined $\Sigma_2^{\mathcal{P}} = \I\Sigma_1^{\mathcal{P}}\Sigma_3^{\mathcal{P}}$, we see that 
they are all traceless, hermitian and satisfy the $su(2)$ algebra relation $[\Sigma_n^{\mathcal{P}},\Sigma_m^{\mathcal{P}}] = 2\I\epsilon_{nmk}\Sigma_k^{\mathcal{P}}$. 
From these new generators, we define the $\sun$ (or let say \textit{P-chiralspin}) group transformations as 
\begin{equation}
	\Psi(x)\to\exp(\I\alpha_n\Sigma_n^{\mathcal{P}})\Psi(x),\qquad
	\bar{\Psi}(x)\to\bar{\Psi}(x)\g{4}\exp(-\I\alpha_n\Sigma_n^{\mathcal{P}})\g{4}.
	\label{eq:su2csnew}
\end{equation}

\noindent
where for different parameters $\alpha_n=\{\alpha_1,\alpha_2,\alpha_3\}$, we can get the axial transformations, $U(1)_A$, and $U(1)_P$ transformations in (\ref{eq:u1p2}).

This group is now different from $\su$, but the transformations (\ref{eq:su2csnew}) coincide with the ones in (\ref{eq:su2cs_barpsi}), when we apply them on the spinors $\psi$ and $\bar{\psi}$ calculated in the point $x^{(t)} = (\bm{0},x_4)$. 
Because in this case $\mathcal{P}x^{(t)} = x^{(t)}$ and consequently  $\psi_{-}(x^{(t)}) = 0$ and $\psi_{+}(x^{(t)}) = \psi (x^{(t)})$ by definition (the same apply for $\bar{\psi}_{\pm}(x^{(t)})$). 
Moreover also $U(1)_P \subset\sun$ coincide with $U(1)_4\subset\su$ in the point $x^{(t)}$, since from (\ref{eq:u1p2}), $\psi(\mathcal{P}x^{(t)}) = \psi(x^{(t)})$ and $\bar{\psi}(\mathcal{P}x^{(t)}) = \bar{\psi}(x^{(t)})$. 

However, while $S_F$ is \textit{P-chiralspin} symmetric, the introduction of a gauge interaction in the action $S_{\Int} = \I\INT \bar{\psi}(x)\g{\mu}A_{\mu}(x)\psi(x)$ breaks explicitly $\sun$, in particular its subgroup $U(1)_P$ of (\ref{eq:u1p2}). 
As shown in \cite{Catillo:2021rrq}, gauge configurations with non zero topological charge $Q_{top}\neq 0$ (as instantons) breaks explicitly $\sun$. Hence we need to restrict in the zero topological sector, and in that case a sufficient condition for the gauge field structure is given as $A_4(x) = A_4 (\mathcal{P}x)$ and $A_i(x) = -A_i(\mathcal{P}x)$, for $i=1,2,3$, which makes $S_{\Int}$ invariant under $\sun$. 

Therefore we conclude saying that $\sun$ solves the first two problems which we mentioned at the beginning of this section. The reason is because, since it is a symmetry of the free massless quark action then, it is compatible with the possibility of deconfinement in QCD. 
Nevertheless it remains to see if $\sun$ can explain the same mass degeneration of the \textit{truncated studies}, originally explained by $\su$. We see this point in the next section.

\section{Correlators}
\label{sec:correlators}

As we have done in section \ref{sec:chiralspin}, here we concentrate on mesons, but for baryons the argument does not change much as outlined in Refs. \cite{Catillo:2021awx,Catillo:2021rrq}. 
Besides Eq. (\ref{eq:corr}), another way of getting meson masses is to fix for example the space $\bm{x} = 0$ and consider the correlators

\begin{equation}
	C_X (\bm{0},t) = \langle O_X(\bm{0},t)\bar{O}_X(\bm{0},0)\rangle,
	\label{eq:cx}
\end{equation}

\noindent
with $O_X(\bm{0},t) = \bar{\psi}(\bm{0},t)X\psi(\bm{0},0)$ and $\bar{O}_X(\bm{0},0)=\bar{\psi}(\bm{0},0)\g{4}X^{\dagger}\g{4}\psi(\bm{0},0)$. 
For large $t$, we still have the exponential decay with the meson mass $m_X$, i.e. $	C_X (\bm{0},t) \sim\exp(-m_X t)$ and it can be evaluated by  computation of the quark propagator as in (\ref{eq:corr2}), since we can rewrite (\ref{eq:cx}) as 

\begin{equation}
\begin{split}
C_X (\bm{0},t) &= \langle \Tr(XD^{-1}(\bm{0},t;\bm{0},t))\Tr(\g{4}X^{\dagger}\g{4}D^{-1}(\bm{0},0;\bm{0},0))\\
& - \Tr(D^{-1}(\bm{0},t;\bm{0},0)\g{4}X^{\dagger}\g{4}D^{-1}(\bm{0},0;\bm{0},t)X)\rangle.
\end{split}
	\label{eq:cx2}
\end{equation}

As we have seen in section \ref{sec:chiralspin}, the presence of $\su$ comes from the fact that the $m_X$s for $X \in\{\Gamma,\Gamma\Sigma_1,\Gamma\Sigma_2,\Gamma\Sigma_3\}$ are all equal. 
However we have also observed that $\su$ and $\sun$ transformations when applied on $\psi(\bm{0},t)$ and $\bar{\psi}(\bm{0},t)$, are the same. 

Consequently this also applies on the two observables $O_X(\bm{0},t)$ and $\bar{O}_X(\bm{0},0)$, which transform in the same way under $\su$ and $\sun$. Therefore the correlators $C_X (\bm{0},t)$ in (\ref{eq:cx}) for $X \in\{\Gamma,\Gamma\Sigma_1,\Gamma\Sigma_2,\Gamma\Sigma_3\}$ can be connected by $\su$ or $\sun$. 
We can't distinguish them.
Moreover, since from $C_X (\bm{0},t)$ we can still get the masses $m_X$s, if we see that they are the same, then this can come from $\su$ or $\sun$ symmetry. 
Hence even in this case, we can't distinguish which type of symmetry is responsible for that. 
This is why $\sun $ can be also suitable for explaining the mass degeneration in the \textit{truncated studies}. This line of thought can be easily extended for whatever hadrons, baryons too. 
Repeating the same argument of section \ref{sec:chiralspin} and Refs. \cite{Glozman:2017dfd,Glozman:2016ayd}, we can not therefore exclude that \textit{P-chiralspin} symmetry is present at high temperature QCD and this line of research would deserve more investigation.

We have said in the previous section (and proved in Refs. \cite{Catillo:2021rrq,Catillo:2021awx}) that $\sun$ is a symmetry of the free massless quark action. 
Therefore we expect a degeneration of the correlators $C_X (\bm{0},t)$ for $X \in\{\Gamma,\Gamma\Sigma_1,\Gamma\Sigma_2,\Gamma\Sigma_3\}$ if we calculate them on the quark propagator in the free case, see eq. (\ref{eq:cx2}). 
Let us check this. 
Now the quark propagator for free massless quarks is simply \cite{Novikov:1983gd} $D^{-1}_{free}(x,y) = \g{\mu}(x-y)_{\mu}/[2\pi^2(x-y)^4]$. 
However it has a pole in $(x-y)^2 = 0$, and we regularize it considering a parameter $\epsilon$ which after the calculation of $C_X(\bm{0},t)$ one can take the limit $\epsilon\to 0$, which means considering $D^{-1}_{free}(x,y)^{(\epsilon)} = \g{\mu}(x-y)_{\mu}/[2\pi^2(x-y)^4 + \epsilon]$. 

Taking for example $X=\Gamma$ and inserting $D^{-1}_{free}(x,y)^{(\epsilon)}$ inside Eq. (\ref{eq:cx2}), where in our case $x$ and $y$ can be $(\bm{0},t)$ or $(\bm{0},0)$, then we get that the disconnected term is zero, since in that case $x=y$. Using only the connected part we get

\begin{equation}
	C_{\Gamma} (\bm{0},t)_{free} = -\lim_{\epsilon\to 0}\langle \Tr(D^{-1}_{free}(\bm{0},t;\bm{0},0)^{(\epsilon)}\g{4}\Gamma^{\dagger}\g{4}D^{-1}_{free}(\bm{0},0;\bm{0},t)^{(\epsilon)}\Gamma)\rangle = \frac{1}{4\pi^4 t^6}\Tr(\Gamma^{\dagger}\Gamma).
	\label{eq:cx3}
\end{equation}

\noindent
As we observe under substitution $\Gamma\to\Gamma\Sigma_n$ with $\Sigma_n$ given in (\ref{eq:su2cs_generator}), $	C_{\Gamma} (\bm{0},t)_{free} $ does not change. 
Thus $	C_{X} (\bm{0},t)_{free} $ for $X \in\{\Gamma,\Gamma\Sigma_1,\Gamma\Sigma_2,\Gamma\Sigma_3\}$ are all equal, because $\Tr(\Gamma^{\dagger}\Gamma) = \Tr((\Sigma_n\Gamma)^{\dagger}\Gamma\Sigma_n)$ for $n=1,2,3$. 
Therefore $\sun$ in the free massless case is a symmetry of the theory, as expected to be from Ref. \cite{Catillo:2021rrq}. 

Let us now move forward. Suppose there is some regime at high temperature where $\sun$ is a symmetry in QCD, then, if we switch on the chemical potential, $\sun$ still remains a symmetry of the theory. 
This simply comes from the fact that the chemical potential term in the action, i.e. $S_{(\mu)}=\mu\INT \bar{\psi}(x)\g{4}\psi(x)$, is $\sun$ invariant.
Indeed from the definition of $\psi_{\pm}$, $\bar{\psi}_{\pm}$ and Eq. (\ref{eq:psi}) we have 

\begin{equation}
	S_{(\mu)}=\mu\INT(\bar{\psi}_{+}(x)\g{4}\psi_{+}(x) +\bar{\psi}_{-}(x)\g{4}\psi_{-}(x)) = \mu\INT\bar{\Psi}(x)\g{4}\Psi(x),
	\label{eq:smu}
\end{equation}

\noindent
where we omitted the terms $\INT\bar{\psi}_{\pm}(x)\g{4}\psi_{\mp}(x)$ because they are zero for parity reasons. 
Now $S_{(\mu)}$ in (\ref{eq:smu}) is of course invariant under $\sun$ transformations given in (\ref{eq:su2csnew}), which is what we wanted to show.

\section{Conclusion}
\label{sec:conclusion}

We have seen that the result of \textit{truncated studies} \cite{Denissenya:2014poa,Denissenya:2014ywa,Denissenya:2015mqa}, namely the large mass degeneration coming from the truncation of the quark propagator (\ref{eq:prop}) which has been explained by the existence of \textit{chiralspin} $\su$ symmetry, can be also described by another group, that we have called $\sun$, or in words \textit{P-chiralspin} group. 

$\sun$, differently from $\su$, is a symmetry of free massless quark action, as chiral and axial group. 
This fact makes $\sun$ compatible with the high temperature regime of QCD, where quarks becomes deconfined. 
Therefore, since lattice QCD results have shown that $\su$ symmetry is present approximately in the range of temperature $T_c - 3T_c$ ($T_c$ temperature of chiral symmetry restoration), we can expect to have $\sun$  at high temperature too. 
If so, we have shown that such \textit{P-chiralspin} persists at non-zero chemical potential, because the chemical potential part of the action is $\sun$ invariant. 
Nevertheless its presence above $T_c$ in QCD is something to be checked in future works.

\section*{Acknowledgements}

	I am grateful to L. Glozman who introduced me to this topic and I am really thankful to M. Marinkovic for the support.

\begin{appendix}

\end{appendix}

\bibliographystyle{SciPost_bibstyle}
\bibliography{chiralspin_catillo_2021_proc.bib}

\nolinenumbers

\end{document}